\documentstyle[12pt,epsf,a4,psfig,cite]{article}
\begin{document}
\baselineskip 0.8 true cm
\def\gms{$G_{Ms}$}
\def\gmp{$G_{Mp}$}
\def\gmn{$G_{Mn}$}
\def\ges{$G_{Es}$}
\def\gep{$G_{Ep}$}
\def\gen{$G_{En}$}
\begin{center}
{\Large\bf Polarization phenomena for meson production in nucleon-nucleon collisions}

\vspace{.2 true cm}

{\rm\large Michail P. Rekalo \footnote{ Permanent address:
\it National Science Center KFTI, 310108 Kharkov, Ukraine} and Egle Tomasi-Gustafsson }

\vspace{.2 true cm}

{\it DAPNIA/SPhN, CEA/Saclay, 91191 Gif-sur-Yvette Cedex,
France}

\end{center}

\begin{center}
{\Large\bf Abstract}
\end{center}

We analyze polarization phenomena for pseudoscalar and vector
mesons production in nucleon-nucleon
collisions. We identify three energy regions corresponding to 
different physics and different approaches in the analysis of polarization
effects. In the threshold
region, characterized by the S-wave production for all final particles, 
the general symmetry properties of strong
interaction can be applied. The
region of intermediate energies, T=2-4 GeV, is characterized by the essential
role of central i.e. non-peripheral collisions, where 
only a small number of $s$-channel states with definite quantum
numbers, ${\cal J}^P=1^-$ and $2^+$, contribute.  At higher energies, T $\ge $10 GeV, the leading mechanism is the diffractive dissociation and it is especially interesting for baryon spectroscopy. The transition to this region is an open field for experimental research at the Nuclotron.

\section{Introduction}     

Meson production in nucleon-nucleon interactions, $N+N-N+N+P(V)$,
where $P(V)$ is pseudoscalar(vector) meson is an important and necessary
ingredient in the study of the nucleon-nucleon (NN) interaction. Our comprehension of "elementary" NN interaction has to be
tested on  meson production. Both the reaction mechanism and the nucleon structure enter in the theoretical model for the 
description of meson production.

Note that from QCD-point of view the processes $N+N-N+N+P(V)$ are
very complicated. No adequate and effective theoretical
scheme in framework of QCD exists, for the description of these processes. But QCD shows some
perspective directions in a more deep understanding of the underlying physics
and models. One example is the importance of non peripheral
central NN-collisions (non-Yukawa picture), when the six-quark intermediate bags play an important role \cite{Fa99}. Such approach seems  very powerful and
predictive. Another example of QCD-inspired problem is $\phi$  and
$\eta$ production in NN-interaction, which can be sensitive to the presence
of $s\overline{s}$-nonperturbative components inside of nucleon.

It is well known that polarization effects
in particle and nuclear physics are very important, because all
fundamental interactions are spin dependent. In particular, for the processes under consideration, spin degrees of freedom play an important role.

Let us raise a very general list of
the main physical problems which could be solved in different
polarization experiments:

\begin{enumerate}
\item  TEST OF SYMMETRY PROPERTIES OF FUNDAMENTAL INTERACTIONS. The
classical example of such test is the experiment of Wu et al
concerning the decay of polarized $^{60}Co$, where  violation of
P-invariance has been discovered. Another example is the test of
CPT-invariance through precise measurements of electron and positron
magnetic moments by comparing the depolarization frequencies of both
leptons in storage rings \cite{pdg}.
\item  EXACT MEASUREMENTS OF FUNDAMENTAL CHARACTERISTICS OF ELEMENTARY
PARTICLES, such as the magnetic moments of electron, proton, neutron, hyperons amd muon (so called g-2 experiment, with very intriguing
results). As an illustration, let us mention the measurement of the proton
electric form factor \gep\  in the scattering of polarized electrons \cite{Jo00}. The
behavior of \gep, which strongly deviates from the dipole parametrization, can be considered as the most surprising result, in the recent years, concerning nucleon structure.
\item  IDENTIFICATION OF REACTION MECHANISM.
\item MULTIPOLE AND PARTIAL WAVE ANALYSIS. Effective multipole analysis
in pseudoscalar and vector meson photo production on nucleons,
$\gamma+N\to N+\pi$, $N+\eta$, $N+\omega$, etc., can be done on the basis of a precise and large amount of data concerning different polarization observables as functions of photon energy and production angle. The same is correct for partial wave analysis of data
about $\pi N$- and $KN$-scattering. All modern physics of nucleonic and
hyperonic resonances is based on these analysis \cite{pdg}.
\item POLARIMETRY OF HIGH ENERGY PROTON BEAM can be done through the
scattering by polarized atomic electrons or through pion production in
$p+Z\to p+Z+\pi^0$, the Primakoff effect \cite{Ca90}.

\end{enumerate}

These problems can be considered as essential points of a polarization programme which can be suggested for
pseudoscalar and vector meson production in NN-collisions:
$p+p\to p+p+P$, $P=\pi$, $\eta$, $\eta^\prime$, $p+p\to p+p+V$, 
$V=\rho$, $\omega$, $\phi$, $p+p\to p+K+\Lambda$, etc.

In order to explain this in more detail, let us introduce three different energy regions for these processes. This classification is based on physical reasons: it follows the changing in the reaction
mechanism, and of the theoretical formalism which is best adapted to its description. Let us indicate the following kinematical regions for
meson production in terms of the kinetic energy of proton beam in the laboratory (LAB) system, $T$:
\begin{itemize} 

\item THE THRESHOLD REGION, i.e. $ E_{thr}\le T \le E_{thr}+\Delta E$ where $E_{thr}$ is the threshold energy and $\Delta E$ depends strongly on the considered reaction.
\item THE INTERMEDIATE ENERGY, $T \simeq $ 2 $\div$ 4 GeV.
\item  THE HIGH ENERGY REGION, or the region of diffractive dissociation (DD), $T \ge $10 GeV.
\end{itemize} 
 
Comparing this classification with the previous list of polarization
problems we note that the study of the three regions is subordinated to the identification of the reaction mechanism for meson production, with the help of  different polarization observables.

We will show later how the diffractive dissociation of high
energy protons and hyperons can be a useful tool to study baryon
spectroscopy, with subsequent test of symmetry predictions, which
follow from quark models, SU(6)-symmetry.. In the frame of the photon-Pomeron analogy, DD can be
considered as a complementary and efficient way for the study of the electromagnetic characteristics of baryons, such as the amplitudes of
radiative decays $Y^*\to Y+\gamma$, $N^*\to N+\gamma$ and the corresponding
electromagnetic form factors. DD could be used also for the
production of polarized  beams of nucleons and hyperons and for
polarimetry of high energy  baryon and anti-baryon beams.

The threshold region can be interesting for non-standard physics, such
as, for example, the test of Pauli principle for identical protons - at
high energies.

Detailed discussion and examples on the ideas listed above will be the object of this report.

\section{The threshold energy region}

\begin{figure}
\label{fig:thr1}
\end{figure}

For three particles production processes like meson production in NN-collisions, the threshold region can be exactly defined in terms of  the orbital angular momenta of the NN-system $\ell_1$ and of the meson $P(V)$, $\ell_2$ (Fig. 1), as the region where
\begin{equation}
\ell_1=\ell_2=0
\label{eq:thr}
\end {equation}

The condition (\ref{eq:thr}) is valid in an energy interval beyond the threshold point, $E=E_{thr}+\Delta E$, where the width of this interval depends on the considered reaction; In any case, $\Delta E\ne 0$, due to the radius of the strong interaction, which is responsible of the process under consideration. In the case of production of open strangeness ($p+p\to p+\Lambda(\Sigma)+K$) or hidden strangeness ($p+p\to p+p+\phi$) this radius is smaller than for $\pi$ or $\rho$ production, making the domain of validity of Eq. (\ref{eq:thr}) larger for strange particle production. 
This definition of threshold region can be experimentally tested by measuring different angular distributions of the produced particles; moreover the T-odd polarization observables, such as analyzing powers or polarizations of the final baryons are very sensitive to even small contributions with non-zero $\ell_1$ and $\ell_2$.
The threshold region has specific properties \cite{Re97a,Re97b,Re97c}. First of all, the essential simplification of the matrix elements induces a simplification of the polarization phenomena. This is related to the presence of a single physical direction, given by the beam momentum $\vec k$, because the S-production, in the final state, corresponds to angular isotropy of the final particles.

\vspace*{-1 true cm}
\begin{minipage}[t]{7 cm}
\mbox{\psfig{file=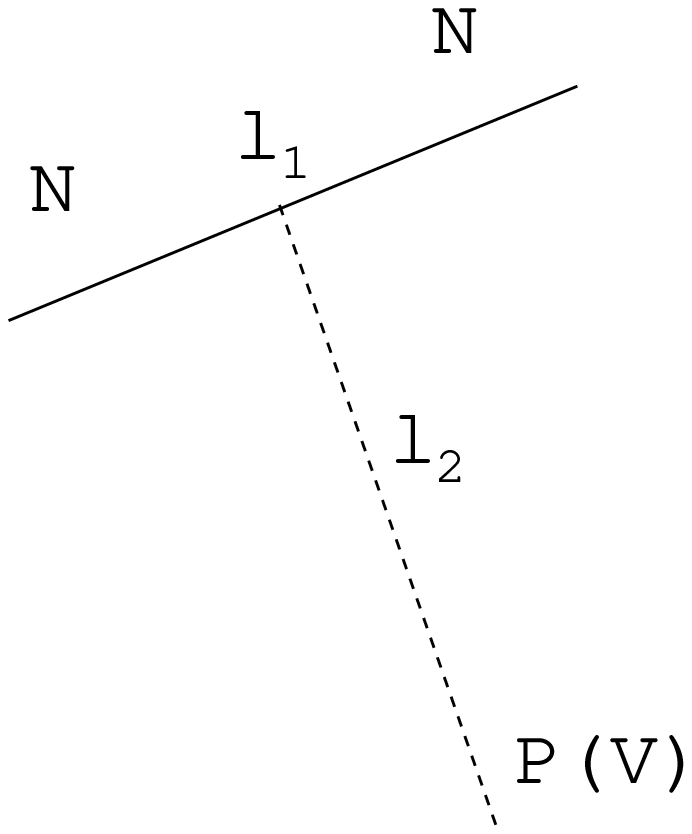,height=6 cm,width=6cm}}\\
\end{minipage}

\vspace*{-4 true cm}
\begin{flushright}
\begin{minipage}[b]{8 cm}
{\bf Fig. 1} Definition of orbital angular momentum for a three-particle system.
\end{minipage}
\end{flushright}

\vspace*{1 true cm}

No reaction plane can be defined, and the reaction is characterized by azimuthal symmetry of the final distributions. Another important point is that symmetry properties of the strong interaction, based on Pauli principle, P-invariance, and  conservation of total angular momentum, strongly constrain and simplify the spin structure of the matrix element.

The formalism of two-component spinors seems well adapted for describing polarization properties of baryons in initial and final states. It is possible, in this framework, to give a transparent parametrization of the spin structure of the matrix elements, in terms of a limited number of partial amplitudes. These amplitudes describe the  possible transitions between the intial and final states which are defined by a set of quantum numbers as: the total spin of the NN-system, the orbital angular momentum of the colliding nucleons, the total angular momentum, ${\cal J}$, and the P-parity of the entrance, i.e. $s$-channel. The resulting spin structure, which corresponds to $s$-channel considerations, well applies to the processes in the threshold region. Note, in this connection, that $t-$channel contributions, which correspond to different mesons, lead to a different parametrization of the spin structure. These two equivalent parametrizations can be transformed to each other by the Fierz transformation. In threshold conditions $t-$channel exchange is not the  most probable mechanism. Indeed, at the threshold of $p+p\to p+p+V$  the value of $t$ is very large and can be calculated as $|t|\simeq m_V^2\gg m_\pi^2$. Therefore the contribution of one pion exchange
can not be essential,  because the corresponding pole is very far from the physical region. Therefore many other $t$-exchanges
must contribute coherently, to produce a definite $s$-channel amplitude - with definite symmetry properties. This  is another justification of the importance of the $s$-channel parametrization of the matrix elements in the threshold region.

As an example let us consider polarization effects in $N+N\to N+N+V$ for the threshold region. In non-coplanar kinematics, generally the spin structure is described by 48 complex scalar amplitudes. In coplanar kinematics, at zero azimuthal angle, the number of amplitudes reduces to 24. The particular case of collinear kinematics, where all three-momenta of the final particles are along the beam direction, is described by a set of 7 amplitudes. At threshold, a unique amplitude describes vector meson production, in case of $pp$-collisions. Let us illustrate this, considering the possible quantum numbers in the initial and final state, in $p+p\to p+p+V^0$, in the threshold region.

In the final state of the processes $p+p \rightarrow p+p+V^0$, taking into
account the identity of the two produced protons (Pauli principle), the
$pp$-system can be produced only in the singlet state, therefore there is
only one possible configuration for the total angular momentum ${\cal J}$ and
the parity $ P$, that is ${\cal J}^P=1^- $. In the initial state, 
due to $ P$-parity conservation, only
odd values for the orbital angular momentum $L$ are allowed. As the total
wave function has to be antisymmetric, the two colliding protons have
to be in a triplet state, $S_i=1$. Therefore only the transition: $L=1,$ $S_i(pp)=1$ $\rightarrow$ ${\cal J}^{P}=1^- $ can take place at threshold for the reaction $p+p \rightarrow p+p+V^0$, with matrix element:
\begin{equation}
{\cal M}=g_1(\tilde {\chi}_2~\sigma_y \vec{\sigma}\cdot\vec{k} \times\vec
U^*\chi_1)~(\chi^{\dagger}_4 \sigma_y\ \tilde {\chi}^{\dagger}_3 )\label{eq:vm7} ,
\end{equation}
\noindent where  $\vec U$ is the
3-vector polarization of the produced vector meson and $g_1$ is the complex
amplitude corresponding to the triplet interaction of the colliding
particles.  The formula (\ref{eq:vm7}) is universal in the sense that it is valid for
any reaction mechanism which conserves the $P$-parity and does not
contradict the Pauli principle.

The most important consequence that follows from (\ref{eq:vm7}) is that the matrix       
element of such a complicated process as $p+p \rightarrow p+p+V^0$ is
defined by a single amplitude $g_1$. All the dynamics of the process is
contained in this amplitude and can be calculated in the framework of a
definite model. But the spin structure of the total amplitude is established
exactly by Eq. (\ref{eq:vm7}) in terms of the 2-component spinors and the vector
polarization $\vec U$. Therefore the polarization effects for any reaction
$p+p \rightarrow p+p+V^0$ can be predicted exactly since they do
not depend on the specific form of the single amplitude $g_1$.  
Of course, $g_1$ depends on the nature of the produced meson and in general
$g_1^{\rho}\neq g_1^{\omega}\neq g_1^{\phi} $, so that the differential cross 
section for the different  
$p+p \rightarrow p+p+V^0$ processes may be different, 
but the polarization observables {\it must be same, independently of the type of vector meson produced}.
 
Let us illustrate this in the calculation of the spin correlation coefficients 
in
the reaction $\vec p +\vec p \rightarrow p+p+V^0$, where both protons in the
entrance channel are polarized:
\begin{equation}
\sigma(\vec P_1, \vec P_2)=\sigma_0(1+\hat{\vec k}\cdot\vec P_1 ~
\hat{\vec k}\cdot\vec P_2)\label{eq:vm8}.
\end{equation}

It is easy to see that  the corresponding
correlation parameter is maximal and equal to $+1$. This correlation
parameter does not contain any information about the dynamics of the
considered processes, because Eq. (\ref{eq:vm8}) is directly derived 
from the $P$-invariance of the strong interaction and from the Pauli principle.  

From (\ref{eq:vm7}), it follows that the 
$V^0-$meson can be polarized even in the collision of
unpolarized protons: $\rho_{xx}=\rho_{yy}=\frac
{1}{2}$, $\rho_{zz}=0$, when the $z-$axis is along the initial momentum
direction. Moreover the decay $V^0 \rightarrow \ell^+\ell^-$ (due to the standard
one-photon mechanism) follows the angular distribution:
$W(\theta) \approx 1+cos^2 \theta \label{eq:vm9}$, where $\theta$ is the angle between $\vec k$ and the direction of
the momentum of one of the leptons (in the system where the $V^0-$meson is
at rest). Here we should emphasize that, at threshold, this $\theta$-distribution is universal and does not depend on assumptions of any
definite mechanism of the process $p+p \rightarrow p+p+V^0$, 
as it was predicted in \cite{El95}, where a similar distribution was obtained 
through the vector current $\overline{s} \gamma_{\mu} s$ acting between $s\overline{s}-$pairs in the proton.

Similarly, for the decays $\phi \rightarrow K +\overline{K}$ and $\rho^0
\rightarrow \pi^+ + \pi^-$, the angular distribution of the produced meson
follows a $\sin^2 \phi-$dependence, where $\phi$ is the angle between the
3-momentum of the pseudoscalar meson (in the system where the $V^0$ is at
rest) and the direction of the momentum of the colliding particles.

The study of polarization effects in  $n+p\rightarrow
n+p+V^0$ is more complicated in comparison with the reaction
$p+p\rightarrow p+p+V^0$. Moreover, for $np$-collisions it is necessary to
treat separately the production of isoscalar ($\omega$ and $\phi$) and
isovector ($\rho^0$) mesons.  This is due to the different isotopic
structure of the amplitudes of the processes $ n+p \rightarrow
n+p+\omega(\phi)$ and $p+p \rightarrow p+p+\omega(\phi)$. 

Due to the isotopic invariance in the strong interactions, the spin structure of 
the amplitudes of the process $ n+p \rightarrow n+p+V^0$ with $I=1$ is described 
by Eq. (\ref{eq:vm7}). For $I=0$, if the final $np$-state is
produced in the $S-$state, then the usual total spin of this system must be
equal to 1 (to satisfy the so-called generalized Pauli principle). This
means that the produced $n+p+V^0$-system can have three values
of $ {\cal J}^{P}$: ${\cal J}^{P}=0^-,~1^-\mbox{~and } 2^-$.

From $P$-invariance, only odd values of the angular momentum $L$ are allowed for
the initial $np-$system: $L=1,~3,....$. One can then conclude that this system 
must be in the singlet state,
$S_i(np)=0$. And, finally, the conservation of the total angular momentum
results in a single possibility, namely: 
$S_i(np)=0,~L=1~\rightarrow~{\cal J^ P}=1^-$, with the following matrix element ${\cal M}_0$:
\begin{equation}
{\cal M}_0=\frac{1}{2}g_0(\tilde {\chi}_2~\sigma_y \chi_1) (\chi^{\dagger}_4
\vec\sigma\times\vec U^*\cdot\vec k\sigma_y \tilde
{\chi}_3^\dagger)\label{eq:vm11} ,
\end{equation}
\noindent where $g_0$ is the amplitude of the process $n+p\rightarrow
n+p+V^0$, which corresponds to $np-$interaction in the initial singlet state.

So, the process $n+p \rightarrow n+p+\omega(\phi)$ is characterized by two
amplitudes, namely $g_0$ and $g_1$. One can see easily that these amplitudes
do not interfere in the differential cross-section of the process $n+p
\rightarrow n+p+V^0$ (with all unpolarized particles in the initial and
final states). Therefore we can obtain the following simple formula for the
ratio of the total cross sections:
\begin{equation}
{\cal R}=\frac{\sigma(p+p \rightarrow p+p+V^0) } {\sigma(n+p \rightarrow
n+p+V^0) } =\frac{2|g_1|^2}{|g_1|^2+|g_0|^2} \label{eq:vm12}.
\end{equation}
In the threshold (or near-threshold) region, this ratio is
limited by:  $0\leq{\cal R}\leq 2 $.

We will see now that the ratio ${\cal R}$ (of unpolarized cross sections)
 contains interesting information on a set of polarization observables for the reaction $n+p\rightarrow
n+p+V^0$.  For example, ${\cal A}_1$ and ${\cal A}_2$ are two independent spin
correlation coefficients, defined only by the moduli square of the
amplitudes $g_0$ and
$g_1$:
\begin{equation}
{\cal A}_1=-\displaystyle\frac{|g_0|^2}{|g_0|^2+|g_1|^2}=-1+\displaystyle\frac{\cal R}{2},~~{\cal
A}_2=\displaystyle\frac{|g_1|^2}{|g_0|^2+|g_1|^2}=\displaystyle\frac{\cal R}{2}, \label{eq:vm13}
\end{equation}
i.e.
\begin {equation}
\begin{array} {rcl}
 0~(g_0=0) & \leq -{\cal A}_1 & \leq 1~(g_1=0),\\
 0~(g_1=0) & \leq{~\cal A}_2 & \leq 1~(g_0=0).
\end{array}
\end{equation} 
But the elements of the density matrix of the $V^0$-mesons, produced in
$n+p\rightarrow n+p+V^0$, are independent from the relative values of the
amplitudes $g_0$ and $g_1$ : $\rho_{xx}=\rho_{yy}=\frac
{1}{2}$, $\rho_{zz}=0$. 

The interference of the amplitudes $g_0$ and $g_1$
appears only in the polarization transfer from the initial to the final
nucleons:$
K_x^{x'}=K_y^{y'}=-2{\cal R}e g_0 g_1^*/(|g_0|^2+|g_1|^2).$

Returning now to the process $n+p \to n+p+\phi$ in connection with
the problem of the $s\overline{s}$-component in the nucleon one can mention that 
a measurement of the ratio of cross sections for $p+p\to p+p+\phi$ 
and $n+p\rightarrow n+p+\phi$,
which are directly related to the relative value of the singlet and triplet
amplitudes would allow to measure the  ratio 
$\displaystyle{\frac{|g_0|^2}{|g_1|^2}}$ and confirm the predicted 
$\phi$-production enhancement from the triplet state
of the $NN$-system. Additional information can be obtained from the measurement 
of spin transfer between the initial and final nucleons.
Similarly, it is possible to study different processes of pseudoscalar meson production: $N+N\to N+N+\pi$, $N+N\to N+N+\eta$, $N+N\to N+Y+K$, with $Y=\Lambda$ or $\Sigma$.

Note, in this connection, that it is possible to measure the relative P-parity of the $K$-meson, $P(YNK)$, through the study of polarization phenomena in $K$-meson production \cite{Pak99}.
\section{The region of intermediate energy}

The different processes of pseudoscalar and vector meson production in NN-collisions, in this energy region,  can be globally described under assumption of an underlying mechanism. Following the standard Yukawa description of these processes in terms of $t-$exchange (OBE), one has to take into account isoscalar ($\eta,~\omega)$ and isovector ($\pi,~\rho$) mesons. The interference between such contributions is important to explain the difference between $pp$ and $pn$ collisions. Such study should give a better insight into the meson-exchange picture of baryon-baryon interaction and meson production. It should determine, in particular, meson-nucleon coupling constants, meson-nucleon final state interaction, and define, in general, the reaction mechanism.

In such model, as meson-nucleon coupling constants are not strongly constrained,  and different parametrizations of form factors are possible,  the energy and angular behavior of the differential cross sections can be reasonably reproduced. However such approach fails in the description of the T-odd polarization observables; which are experimentally found to be large, but they vanish in OBE models. There are many other experimental indications about the importance of central collisions: $s-$channel contributions,  with definite ${\cal J}^P$, where ${\cal J}$ and $P$ are the total angular momentum and parity in the $s-$channel, have to be taken into account, in the intermediate energy region. From a theoretical point of view, such contributions are justified in a quark model, \cite{Fa99}, considering six quarks intermediate states, in presence of $\pi$, $\sigma$ and $\rho-$mesons (Fig. 2). These $s$-contributions can produce pole-like amplitudes, with corresponding $s-$singularities in the physical region, whereas $t-$channel contributions essentially decrease at large angles, (i.e. at large $|t|$) because the corresponding $t-$singularities move away from the physical region. The $s-$channel singularities are $t-$independent and therefore equally present at any momentum transfer.  Most of the existing data on the angular and energy dependence of the differential cross sections and analyzing power, for the processes: $p+p\to p+p+\eta$, $p+p\to p+p+\omega$, $p+p\to \Delta^{++} +n$, $p+p\to \Delta^{++} +\Delta^0$ ... can be explained in terms of only two intermediate states, 
${\cal J}^P=1^-$ and ${\cal J}^P=2^+$. 
\vspace*{-1 true cm}
\begin{minipage}[t]{9 cm}
\mbox{\psfig{file=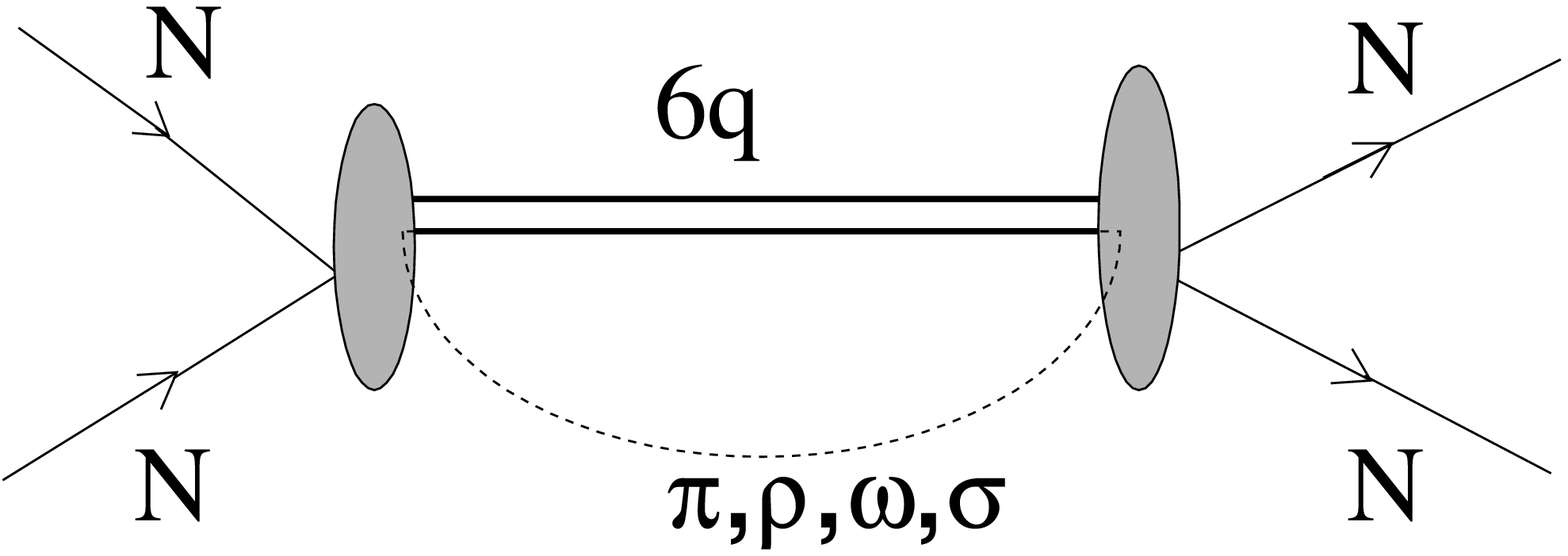,height=5 cm,width=9cm}}\\
\end{minipage}
\vspace*{-3 true cm}
\begin{flushright}
\begin{minipage}[b]{6 cm}
{\bf Fig. 2} Feynman diagrams for six-quarks or meson exchange.
\end{minipage}
\end{flushright}

\begin{figure}
\label{fig:fig2}
\end{figure}
\vspace*{2 true cm}


As an example, let us consider the reaction $p+p\to p+p+\eta$. The standard description is based on the subprocess $p+p\to N^*(1535)+p\to p+p+\eta$, which is realized, in the Yukawa model, through different $t-$exchanges (Fig. 3). It is generally admitted \cite{Pak99,Fa90,La91,Ve91,Gr00} that the $\eta$-production takes place through $N^*(1535)$, since this resonance has a large branching ratio in the $N\eta$-channel. The decay $N^*(1535)\to N\eta$ occurs, with BR=(30-55)\%. The FSI interaction is especially important for the description of T-odd polarization observables. But in the model of central collisions, another interpretation is given, where two $NN$-states, with ${\cal J}^P=1^-$ and ${\cal J}^P=2^+$ play an important role.

To illustrate this, let us consider two different processes, $p+p\to p+p+\eta$ and $p+p\to p+p+\omega$ and  let us compare the corresponding predictions with the existing data on cross sections and analyzing powers.

\begin{flushleft}
\vspace*{-1 true cm}
\begin{minipage}[t]{9 cm}
\mbox{\psfig{file=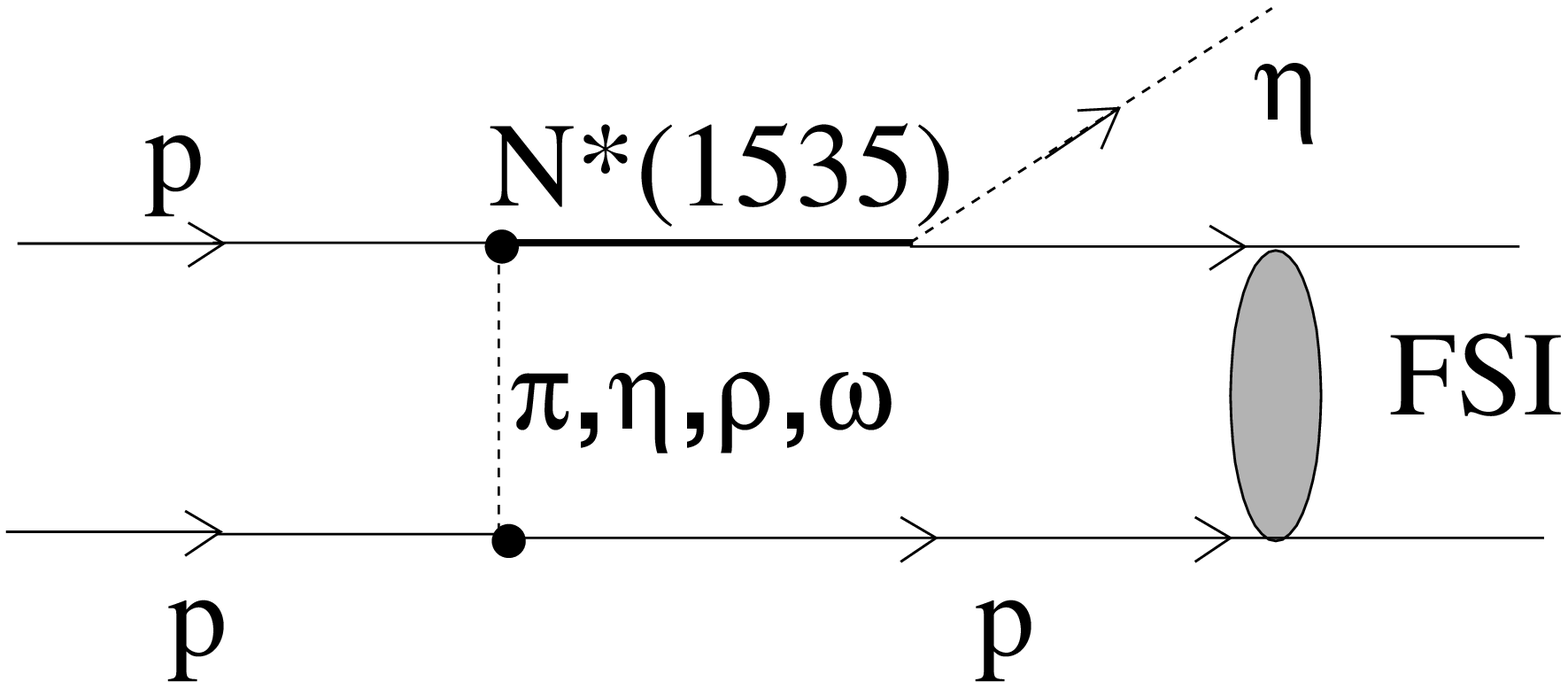,height=5 cm,width=9cm}}
\end{minipage}
\end{flushleft}

\vspace*{-4true cm}
\begin{flushright}

\begin{minipage}[b]{6 cm}
{\bf Fig. 3} Feynman diagrams for $\eta$-production through $t-$channel meson-exchange
\end{minipage}
\end{flushright}

\begin{figure}
\label{fig:fig3}
\end{figure}
\vspace*{1 true cm}


\subsection{$p+p\to p+p+\eta$: small excitation energy for the $pp$-system}

At small excitation energy, $M_{pp}-2m_p\le$ 5 MeV, where $M_{pp}$ is the effective mass of the two protons in the final state, this $pp$-system must be in $^1S_0$-state. Following the Pauli principle, the conservation of the $P$-parity and total angular momentum, the initial state has to be triplet, with odd orbital angular momentum. Therefore $\ell_\eta$ has to be even, and the allowed intermediate $S$-states have ${\cal J}^P=0^-$, $2^-$.. They are forbidden in the central model described above. This is indeed what is found at $T=1520$ MeV \cite{Ta00}. At larger energy, at $T=1805$ MeV, some events are observed as one moves away from the threshold region and other states play a more important role.

\subsection{$p+p\to p+p+\eta$: all $M_{pp}$-events}

Without a particular selection on $M_{pp}$, another situation appears. The final protons can have any value $\ell_1$, but for a limited value of $M_{pp}$ ($M_{pp}-2m_p\le$  40 MeV, \cite{Ta00}), with a number of events rapidly decreasing when $M_{pp}$ increases, we will limit our calculation to $\ell_1=1$. The Pauli principle constrains the value $S_f=1$, where $S_f$ is the spin of the final state, therefore $j_{pp}$, the total angular momentum of the final $pp-$system can take the values 0, 1, 2 - with negative P-parity.
\vspace{.2 true cm}

\underline{Case with ${\cal J}^P=1^-$}. A single transition is allowed here, with the following quantum numbers:
$$ S_i=\ell_i=1 ~\to~ {\cal J}^P=1^- ~\to~ \ell_\eta=1.$$
For simplicity we will consider only one value for $j_{pp}$, $j_{pp}=1$. The corresponding matrix element is parametrized as follows:
$${\cal M}_1=g_1\left (\vec\sigma\times\vec k\right )_a\bigotimes \left ([\vec\sigma\times\vec p ] \right )_a,$$
where $g_1$ is the ${\cal J}^P=1^-$ partial amplitude, $\vec k$ is the unit vector along the beam momentum, $\vec p (\vec q)$ is the unit vector along the three-momentum of the final protons ($\eta$-meson). We are using here the following convention: $$C\bigotimes D =(\tilde \chi_1\sigma_y C\chi_2) (\chi_4^\dagger\sigma_yC\tilde\chi_3^\dagger).$$
Averaging over the colliding proton spins, summing over the final proton spins, and integrating over the $\vec p$ direction, one can find:
$$d\sigma/d\Omega_{\eta}\simeq |g_1|^2(1+\cos^2\theta),$$
where $\theta$ is the angle of $\eta$-meson production.

\underline{Case with ${\cal J}^P=2^+$}. The simplest way to obtain this state, is to consider the following set of quantum numbers:
$$ S_i=0,~\ell_i=2 ~\to~ {\cal J}^P=2^+ ~\to~ j_{pp}=0~(S_f=\ell_f=1),~\ell_2=2.$$
with matrix element:
$${\cal M}_2=g_2 I\bigotimes\vec\sigma\cdot\vec p~k_ak_b(q_aq_b-
\displaystyle\frac{1}{3}\delta_{ab}),$$
which gives the following cross section:
$$d\sigma/d\Omega_{\eta}\simeq |g_2|^2(\cos^2\theta -\displaystyle\frac{1}{3})^2,$$
So, the following angular dependence for the differential cross section of the process $p+p\to p+p+\eta$ (with excitation of non-interfering states ${\cal J}^P=1^-$ and ${\cal J}^P=2^+$) can be written as a linear combination:
$$d\sigma/d\Omega_{\eta}\simeq A(1+\cos^2\theta)+B\left (\cos^2\theta-\displaystyle\frac{1}{3}\right )^2,$$
where $A$ and $B$ can be fitted on the experimental data. A very good agreement is obtained, as shown in Figs. 4 and 5 (from \cite{Ta00}).

\begin{flushleft}
\vspace*{-2 true cm}
\begin{minipage}[t]{7.5 cm}
\mbox{\psfig{file=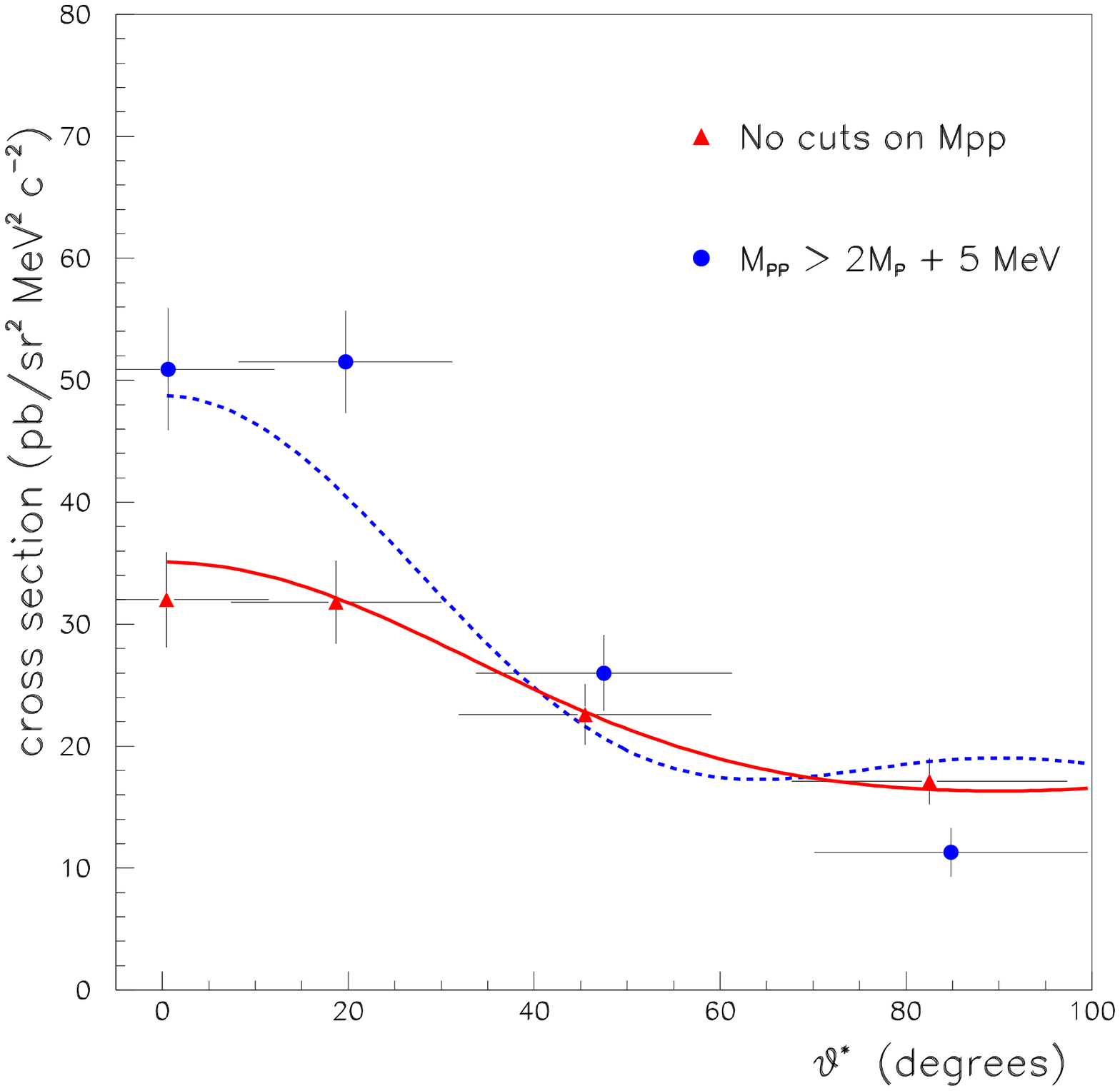,height=8.5 cm,width=8.5 cm}}\\
{{\bf Fig. 4} C.M. cross sections for pp$\to $ pp$\eta$ production at 15205~MeV. Full (dashed) line corresponds to events without cuts on $M_{pp}$
(with cut $M_{pp}\ge 2M_{p}$+5~MeV events).}
\end{minipage}
\end{flushleft}
\begin{figure}
\label{fig:fig4}
\end{figure}

\vspace*{-12 true cm}
\begin{flushright}
\begin{minipage}[b]{7.5 cm}
\mbox{\psfig{file=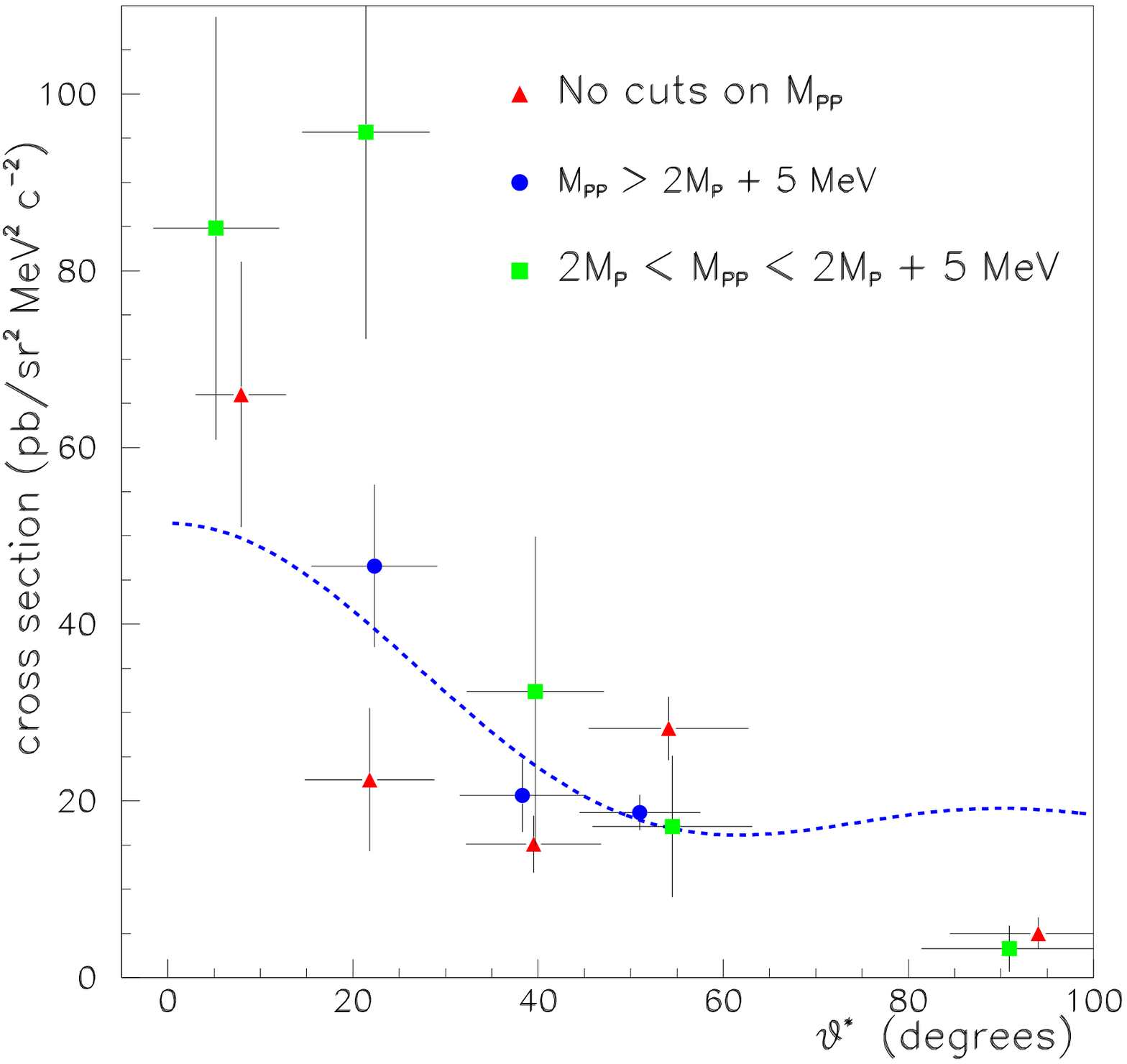,height=8.5 cm,width=8.5cm}}\\
{{\bf Fig. 5} C.M. cross sections for $pp\to pp \eta$ production at 1805~MeV.
The dashed curve corresponds to $M_{pp}\ge 2M _{p}+5$~MeV events.}
\end{minipage}
\end{flushright}

\begin{figure}
\label{fig:fig5}
\end{figure}




\subsection{$p+p\to p+p+\omega$ }

In the same way we can analyze the process $p+p\to p+p+\omega$, which has a more complicated spin structure, due to the presence of a vector meson. Let us consider again the production of $pp$ in $^1S_0$ state. In case of 
${\cal J}^P=2^+$ excitation, the following transition contributes:
$$ S_i=0,~\ell_i=2 ~\to~ {\cal J}^P=2^+ ~\to~ \ell_f=\ell_\omega=1.$$
The spin structure of the corresponding matrix element can be written as:
$${\cal M}_{\omega}=g_sI\bigotimes I(k_ak_b-\displaystyle\frac{1}{3}\delta_{ab})U^*_{\omega} q_b,$$
where $\vec U$ is the polarization vector of the ${\omega}$-meson. This allows to find for the differential cross section:
$$d\sigma/d\Omega_{\omega}\simeq |g_s|^2(1+3\cos^2\theta),$$
again, in very good agreement with the data (Fig. 6) \cite{Ta00}.

\begin{flushleft}
\vspace*{-1 true cm}
\begin{minipage}[t]{11 cm}
\mbox{\psfig{file=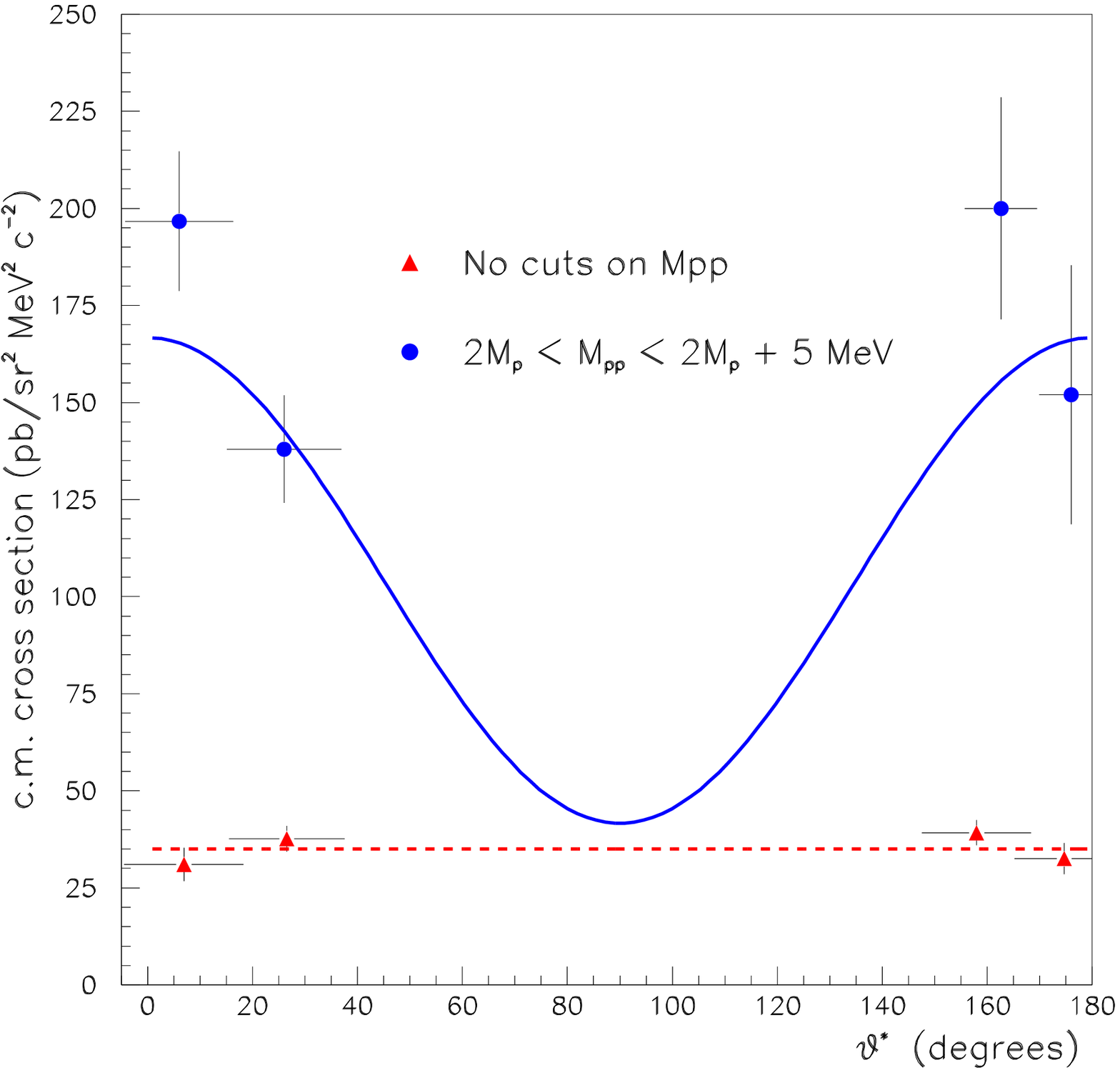,height=10cm,width=10cm}}\\
\end{minipage}
\end{flushleft}

\vspace*{-8 true cm}
\begin{flushright}

\begin{minipage}[b]{6 cm}
{\bf Fig. 6} C.M. differential cross section for pp$\to $ pp$\omega$ production
  at 2100~MeV shown as a function of the $\omega$ c.m. production
  angle. The
  full (dashed) line corresponds to invariant masses of
  $2M_{p}~\le~M_{pp}~\le~2M_{p}~+~5$~MeV
  (without cuts on $M_{pp}$) events.
\end{minipage}
\end{flushright}

\begin{figure}
\label{fig:fig6}
\end{figure}
\vspace*{2 true cm}


Let us consider now the $\omega$-production for all events in $M_{pp}$. As previously discussed, let us consider $\ell_1=S_f=1$, for ${\cal J}^P=2^+$ excitation, with the following matrix element:
$${\cal M}_2^{(\omega)}=g_2(\vec\sigma\cdot \vec p\times \vec k~ \vec k\cdot \vec U^* -\displaystyle\frac{1}{3}\vec\sigma\cdot \vec p\times \vec U^*).$$
After integration over $\vec p$, the differential cross section is isotropic. The same result holds for ${\cal J}^P=1^-$, and it is in agreement with experiment (see Fig. 6).

It is possible to analyze in a similar way other inelastic NN-processes: $p+p\to\Delta^{++}+n$
or $p+p\to\Delta^{++}+\Delta^0$, with very complicated spin structure of the corresponding matrix element. For example, $\Delta\Delta$-production is described by 32 independent complex amplitudes. Models based on different $t$-exchanges exist \cite{Kl80,Mi93,Au89,Ha94}, but, again, they can not reproduce T-odd spin observables, because all $t-$amplitudes are real functions of $t$. The model of central collisions give  naturally complex amplitudes and sizeable values of analyzing powers $ {\cal A}$ result from the interference of ${\cal J}^P=1^-$ and $2^+$ states. More exactly $ {\cal A}\simeq Im f_1f_2^*=|f_1||f_2|\sin\delta$, where $\delta$ is the relative phase of the amplitudes. The analyzing power is determined by two energy-dependent parameters, $\delta$ and $r=|f_1|/|f_2|$. The success of this model is confirmed by the quality of the corresponding fit to the angular dependence of the analyzing powers at different energies (Figs. 7 and 8)\cite{Yo98}. 

\begin{flushleft}
\vspace*{-3true cm}
\begin{minipage}[t]{12 cm}
\mbox{\psfig{file=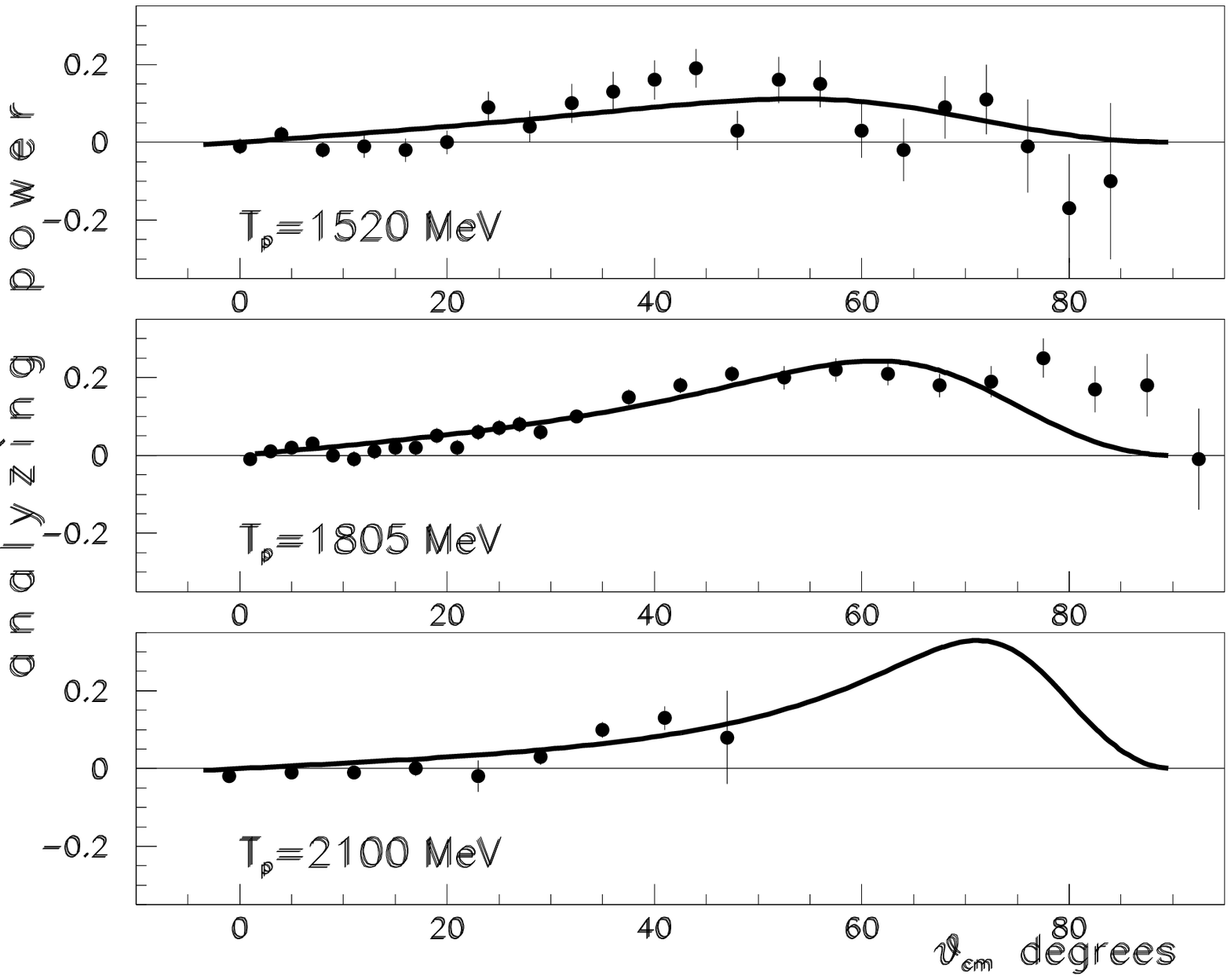,height=12cm,width=12cm}}\\

\end{minipage}
\end{flushleft}

\vspace*{-8 true cm}
\begin{flushright}

\begin{minipage}[b]{4 cm}
{\bf Fig. 7} Analyzing powers at all three energies with and without  $\Delta$ formation
in the region of $ \vec p+p \to \Delta^{++}+\Delta^{0}$ reaction. The curves are theoretical fits discussed in the text.
\end{minipage}
\end{flushright}

\begin{figure}
\label{fig:fig7}
\end{figure}
\vspace*{1.5 true cm}


However this approach does not predict the energy dependence of the phase. But from a physical point of view, we can expect that this phase is a more or less universal function for the different processes, as it can be proved in case of Breit-Wigner parametrization of the considered central partial wave amplitudes. This phase can be calculated in  microscopic descriptions of central collisions \cite{Fa99} and should be experimentally tested.

Note that the model of central collisions  can be applied to different $\gamma d$-processes:
$\gamma+d\to n+p$, $\gamma+d\to d+\pi^0$, $\gamma+d\to \Delta +N$, and to the corresponding electroproduction processes. The corresponding value of the Mandelstam variable $s$ is obtained for $E_\gamma =0.5$ T, therefore the model of central collisions should aply in the energy range $1 \le E_\gamma\le $2 GeV. The differential cross section should be described by a polynome in $\cos\theta$, essentially from an exponential $t$-dependence, typical for impulse approximation. Polarization phenomena should be large. This is a good physics case for the Jefferson Laboratory.

\begin{flushleft}
\begin{minipage}[t]{10 cm}
\mbox{\psfig{file=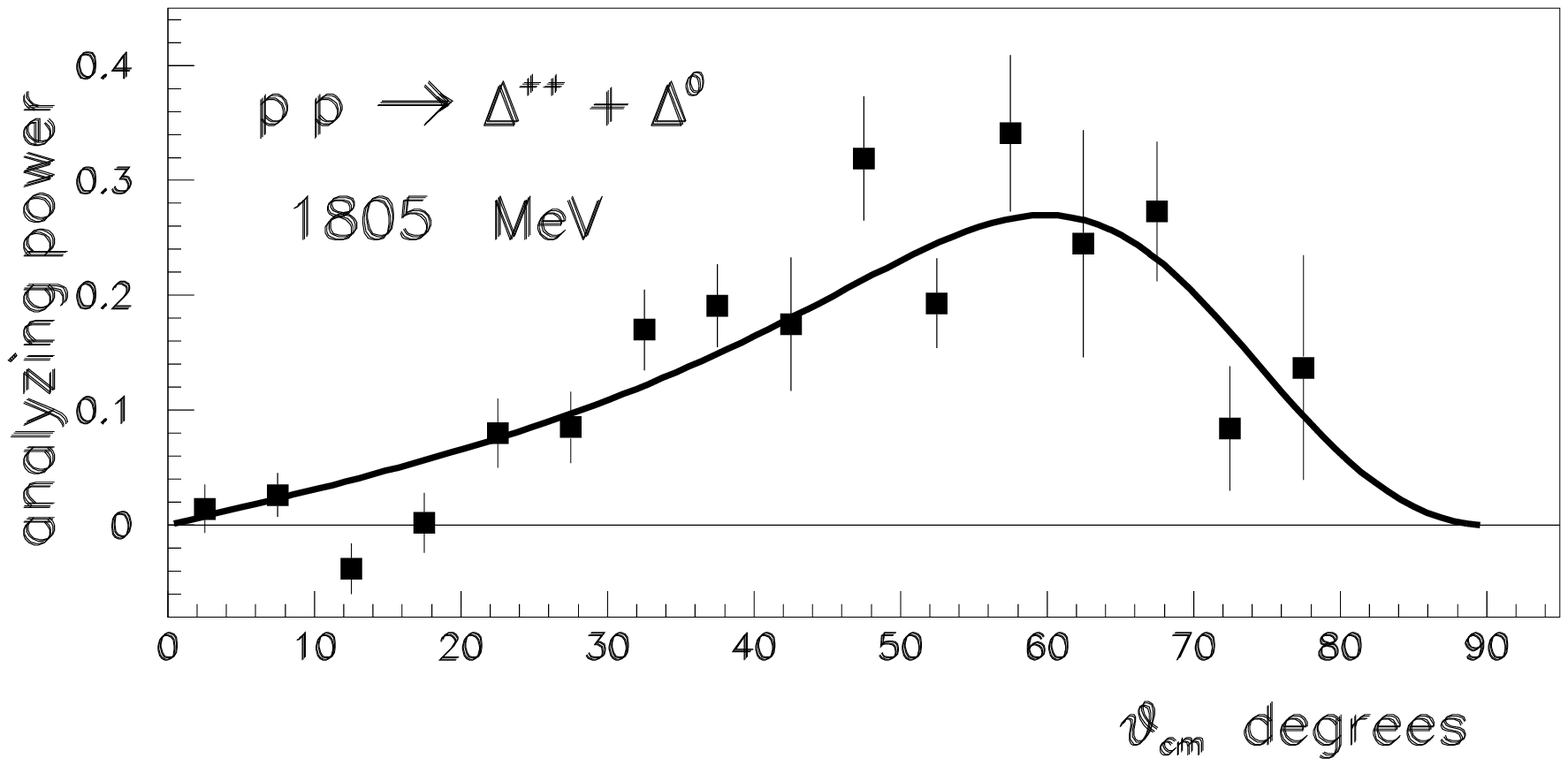,height=10cm,width=10cm}}\\
\end{minipage}
\end{flushleft}

\vspace*{-10 true cm}
\begin{flushright}

\begin{minipage}[b]{6 cm}
{\bf Fig. 8} Analyzing powers with two $\Delta$ 
formation after background subtraction. The curve 
is the theoretical fit discussed in the text.
\end{minipage}
\end{flushright}

\begin{figure}
\label{fig:fig8}
\end{figure}
\vspace*{1.5 true cm}


\section{Region of diffractive dissociation}
In 1953 two russian scientists, I. Pomeranchuk and E. Feinberg, predicted a new type of processes, the inelastic diffractive scattering of high energy hadrons, or diffractive dissociation (DD) \cite{Po53}. Diffraction is a very general phenomenum in different physical regions:
\begin{itemize}
\item classical optics
\item nuclear interaction
\item hadron physics
\end{itemize} 
For diffractive phenomena to occur, two important ingredients are necessary:
\begin{itemize}
\item large value of absorption cross section
\item small wavelength in comparison with the size of the target particles
\end{itemize} 
The possibility of DD has been based on analogy with the well-known 
QED radiation: {\it " It is usually considered that diffraction manifests itself only as elastic scattering. However the change of the motion of a charge, induced by scattering, gives rise to gamma rays. it is evident that such diffraction scattering of nuclear active particles, (nucleons, pions) have to be accompanied by the emission of pions and, possibly, nucleon pairs as well."}
DD can therefore be cosidered as a generalization of the analogy between hadron elastic scattering and classical diffraction, and signs a transition from classical fields (optics) to fields describing relativistic particles: the quantum field nature of particles.

Therefore DD in hadron physics is related to the possible excitation of the internal degrees of freedom during the scattering. In this respect, DD has no analogy with the diffraction of classical waves.

In the middle of the 1960's, with the advent of the high energy accelerators, DD has been observed in experiments with proton and pion beams. Subsequent studies showed that the diffractive mechanism is one of the leading processes for particle generation, essentially contributing to the total cross section for hadron interaction at high energies. DD is important in electromagnetic interactions and in weak interactions, as well.

In modern understanding DD is interpreted as a result of the exchange of a particle ${\cal P}$, named Pomeron, in the memory of I. Y. Pomeranchuk.

Generally, this process (in non-coplanar kinematics) is described by a set of five independent kinematical variables, namely: $s=(p_1+p_2)^2$ is the square of the total energy of colliding particles, $t=(p_2-p_4)^2$ is the momentum transfer squared, $\omega^2=(p_1+p_2)^2$ is the square of the effective energy of the system $p+P(V)$,$\cos\theta$ is the cosinus of the production angle $\theta$ for $P(V)$, in the CMS of the subprocess ${\cal P}+p\to N+P(V)$, and $\phi$ is the azimuthal angle which characterizes the acoplanarity of the $p+p\to p+N+P(V)$, i.e. the angle between the plane of the reaction  ${\cal P}+p\to N+P(V)$ and the plane defined by the $\vec p_1$ and $\vec p_4$ three-momenta.

The typical kinematical conditions for the applicability of the Pomeron exchange can be formulated as follows: $s \gg M^2$, $|t|$, $\omega^2$, where $M$ is the nucleon mass. To be more precise, we can also write: $|t|\le $ 1 GeV$^2$, $\omega$ =1$\div$ 2.5 GeV, $T\ge 10$ GeV. There is, in general, no constrain on the other two variables, and we can write:$-1\le\cos\theta \le 1$, $0\le\phi\le 2\pi$.

The matrix element for the Pomeron exchange can be written, neglecting the spin degrees of freedom of the interacting particles:
$${\cal M}=g_{NN{\cal P}}g_{{\cal P} N\to NP(V)}(\omega,t,\cos\theta) s^{\alpha(t)}\displaystyle\frac{1+e^{i\pi\alpha(t)}}{\sin\pi\alpha(t)}$$
where $g_{NN{\cal P}}$ is the corresponding form factor for the $NN{\cal P}-$vertex, $g_{{\cal P} N\to NP(V)}$ is related to the the amplitude for the subprocess ${\cal P}+p\to N+P(V)$, $\alpha(t)$ is the main characteristics of the Pomeron phenomenology, the vacuum trajectory, and ${1+e^{i\pi\alpha(t)}}/{\sin\pi\alpha(t)}$ is the 'signature factor'. The diffraction matrix element does not depend on $\phi$, as we neglected the spins of all particles, including the Pomeron, the quantum numbers of which must be equal to the quantum numbers of vacuum-vacuum exchange.

There is, however a problem, concerning the spin of Pomeron. In the Regge formalism, the trajectory $\alpha(t)$ is playing a role of the corresponding spin. It is the factor $s^{\alpha(t)}$ which drives the very specific energy dependence of the Regge contribution, unifying simultaneously the $s-$ and $t-$ dependences of the amplitudes. However, in order to write the correct matrix element, one has to know the resulting spin structure, and the main problem is the description of the spin properties of the Pomeron.

One possible assumption is that the spin of the Pomeron is equal to 1, like a virtual photon. This is the basis of the so-called {\it photon-Pomeron analogy} \cite{Do83}. Consequently, it is possible to apply the standard technics of the modern theory of pion photo- and electroproduction (in the resonance region) for the description of the diffractive subprocesses:${\cal P}+p\to N+\pi$, $N+\eta$, $N+V$ etc..

Let us remind the three typical contributions which are important for the description of the reactions $\gamma+N\to N\pi$.
\begin{itemize}
\item Born contributions
\item vector and axial meson exchange
\item $N^*$-excitations
\end{itemize} 
Similar mechanisms can be applied to ${\cal P}+p\to N+\pi$, (Fig. 9). A large amount of data about differential cross sections and polarization phenomena on the subprocess ${\cal P}+N\to B+P(V)$, where B is the final baryon, nucleon or hyperon can be discussed in terms of partial wave analysis, and properties of $N^*$ and $Y^*$-resonances. Data exist on  $N^*$ and $Y^*$ produced in general, througn $\pi N$ and $\overline{K}N$-formation experiments, however importan questions are still open. For example, quark models predict that some excited states couple strongly to $\pi N$ or $\overline{K}N$-channels, whereas other states almost decouple. This is correct, in particular in the $S=1$ sector. The quark model predicts numerous excited $\Lambda$ and $\Sigma$-baryons, which have not been observed. The $\Omega$-sector is almost totally unexplored.

Therefore one possible way to find the missing baryon resonances is to study the channels that couple more strongly to these missing states. In the case of $N^*$ the decays $N^*\to N+V$, $ V=\rho$,$\omega$, $N+\eta$, $N+\pi\pi$ could be investigated with high statistics, and interpreted in the frame of Pomeron exchange.

To illustrate the situation, let us remind that the last editions of the Particla data Group review  \cite{pdg} (during 10 years!) have prefaced the notes
on  $ \Lambda  $ and $ \Sigma  $
resonances with the comment:
{ ..there are no new results at all on $ \Lambda $ and $ \Sigma $ resonances. The field remains at a stand still and can only be revised if a
KAON factory is built. } A possible alterantive to Kaon factory can be the DD of high energy hyperons. 

We mentioned above the similarity of DD to photo(electro)-production processes. Let us stress some important points in this comparison:
\begin{itemize}
\item The photon is a C-odd particle, $C(\gamma)=-1$ , whereas $C({\cal P})=+1$: the Pomeron is a $C$-even object, which interacts equally with particles and antiparticles; therefore, for example, $g({\cal P}\pi^+\pi^+)=g({\cal P}\pi^0\pi^0)$;
\item  $I({\cal P})=0$: the Pomeron can be considered as an isoscalar photon with positive  $C$-parity;
\item $V-$meson exchange is forbidden in ${\cal P}+N\to N+\pi$, the $\Delta$-contributions, for any ${\cal J}^P$ can not be excited. Therefore DD can be considered as an isotopic filter, very efficient for the selection of baryonic resonances.
\end{itemize} 

Let us stress some of the advantages of DD, in comparison with electromagnetic probes:
\begin{itemize}
\item The cross section is at least two order of magnitude larger;
\item Flexibility of baryon targets: one can investigate exotic processes, such as ${\cal P}+\Lambda (\Sigma)\to \pi+Y$, ${\cal P}+\Omega\to Z+K$, which can not be studied in any other way;
\item absence of radiative corrections
\end{itemize} 
The $\gamma{\cal P}$-analogy can be symbolically written as:
$${\cal M}_{DD}=V_{\mu}(1)V_{\mu}(2),$$
where $V_{\mu}(1)$ and $V_{\mu}(2)$ are the vector currents corresponding to the two vertexes of the ${\cal P}$-exchange diagram. Such hypothesis results in a definite $\phi$-dependence of the differential cross section for any process of DD:
\begin{equation}
\displaystyle\frac{d^4\sigma}{dt d\omega d\cos\theta d\phi}=a_0(t,\omega,\cos\theta)+ \cos\phi a_1(t,\omega,\cos\theta)+\cos2\phi a_2(t,\omega,\cos\theta),
\end{equation}
similarly to the electroproduction cross section. No experimental verification has been done up to now.

\mbox{\psfig{file=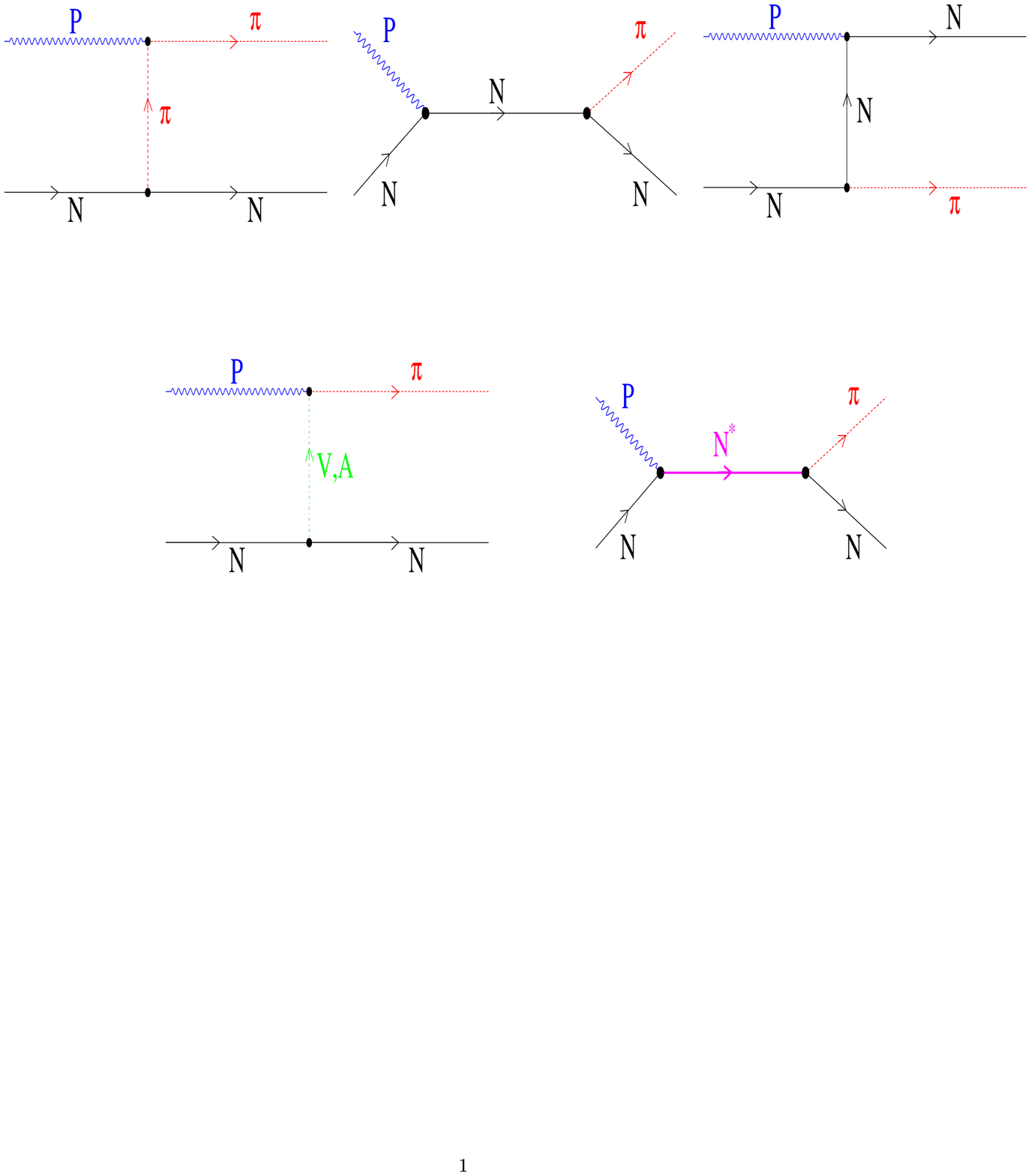,height=10cm,width=15cm}}\\
\vspace*{-1.5truecm}
\begin{center}
{\bf Fig. 9} Feynman diagrams for Pomeron exchange.
\end{center}

\begin{figure}
\label{fig:pom}
\end{figure}
\
Let us summarize the main point developped in this chapter, concerning DD induced by different beams: nucleons, mesons, hyperons etc.:

\begin{itemize}
\item DD is a general phenomenon for strong, electromagnetic and weak interactions, at high energies;
\item DD is characterized by a large cross section, almost $s$-independent, which is an essential part of the total cross section of hadron-hadron interaction;
\item DD is the object of non-perturbative QCD;
\item DD can be viewed as an alternative way to study the physics of baryon resonances, in particular for hyperons.
\end{itemize}

\section{Conclusions}

We have discussed the physics related to meson production and classified the different features with respect to three kinematical regions.

\subsection{The threshold region}

\noindent\underline{The main properties}
\begin{itemize}
\item essential simplification of the spin structure of the threshold matrix elements - due to the general symmetry properties of the strong interaction, such as the Pauli principle, the P-invariance, the isotopic invariance, the $C$-invariance etc.;
\item corresponding simplification of polarization phenomena;
\item direct connection, at the level of polarization observables, between the internal and the space-time symmetry properties of fundamental interactions - for example, the relation of the isotopic invariance with the P-invariance and the conservation of the total angular momentum.

\end{itemize} 

\noindent\underline{What is interesting to measure}
\begin{itemize}
\item Determination of the size of the threshold region for $\eta$, $\eta'$, $K$ and $V$-meson production, through the study of T-odd polarization phenomena, which vanish for $S$-wave production of the final particles;
\item Test of model independent predictions for polarization phenomena;
\item Comparison of $pp$- and $np$-meson production, as the simplest and more direct way to test the physics of the OZI-violation;
\item Determination of the role of $P$-wave production;
\item Partial wave analysis for $\eta$, $\omega$ or $\phi$ production with the help of polarization effects.
\item Test of symmetry properties of the strong interaction (for example, test of the validity of the Pauli principle for high energy protons)
\end{itemize} 

\subsection{The intermediate energy region}

\noindent\underline{The main properties}
\begin{itemize}
\item Essential role of central collisions for meson production (manifastation of non-Yukawa mechanism in $NN-$collisions) - with excitation of few states (six-quark bags?) with definite quantum numbers ${\cal J}^P=1^-$ and ${\cal J}^P=2^+$. 
\item Polynomial (i.e. almost flat, non exponential) $\cos\theta$-behavior of the differential cross sections and different polarization observables
\item Polarization obseravbles as analyzing powers or final baryon polarizations are large in absolute value
\item A relativelely fast change in the shape of the angular dependence of the cross section and of polarization observables, in a relatively small kinematical interval - about 300-400 MeV.
\end{itemize} 
\noindent\underline{What is interesting to measure}
\begin{itemize}
\item The angular dependence of the differential cross sections and of the polarization observables for different non-elastic processes: $N+N\to \Delta +N$, $\to \Delta +\Lambda$, $\to N +N+\eta$,$N +N+V$...
\item Determination of the quantum numbers ${\cal J}^P$ of the intermediate quark bags, through their decays: $6q\to NN\pi$, $NN\eta$, $NNV$;
\item Determination of the spin structure of the matrix elements for these decays.
\end{itemize}

\subsection{The diffractive dissociation}

\noindent\underline{The main properties}
\begin{itemize}
\item Large value of the corresponding cross section;
\item Universality: DD is present in all processes of pseudoscalar and vector meson production in NN-collisions: $ N+N\to N+N+{\cal P}$ and $ N+N\to N+N+V$;
\item Unification of high energy phenomenolgy (i.e. the validity of the Pomeron exchange at large $s$) with non-perturbative low energy physics, which is responsible for the subprocess ${\cal P}+ N\to N+P(V)$;
\item Large polarization phenomena, particularly in exclusive experiments, with detection of at least two final particles;
\item Specific dependence on the azimuthal angle $\phi$, as manifestation of the spin properties of the Pomeron exchange.
\end{itemize} 
\noindent\underline{What is interesting to measure}
\begin{itemize}
\item Differential cross section and polarization observables in specific kinematical conditions, i.e. at small momentum transfer $t$, relatively small excitation energy (in the resonance region) - in order to perform a multipole analysis of processes as ${\cal P}+ N\to N+\eta$, $N+V$ ... in a complementary way with respect to electromagnetic probes; this is also interesting for hyperon diffractive dissociation;
\item Special interest has to be devoted to the question of the spin structure of Pomeron exchange -through the study of different Pomeron-hadron vertexes. 
Elastic hadron-hadron scattering is not convenient, for this purpose, because polarization phenomena are very similar to QED: vertexes are essentially 'real', and the simplest non vanishing observables involve spin correlations.
On the contrary, for DD the one-spin polarization observables are different from zero.
\end{itemize} 

We would like to thank our collaborators in different steps of this work, in particular J. Arvieux, B. Tatischeff and J. Yonnet. We thank the organizers to give us the possibility to be here and to honour the memory of Academician Prof. A. M. Baldin, who did so much for this field of physics.


\begin{thebibliography}{}

\bibitem{Fa99}A. Faessler, V. I. Kukulin, I. T. Obukhovsky and V. N. Pomerantsev, e-print: nucl-th/9912074.
\bibitem{pdg} Particle Data Group, D.E. Groom {\it et al.}, The European Physical Journal C {\bf 15}, 1 (2000).
\bibitem{Jo00} M. K. Jones {\it et al.}, Phys. Rev. Lett. {\bf 84}, 1398 (2000).
\bibitem{Ca90} D. C. Carey {\it et al.}, Phys. Rev. Lett. {\bf 64}, 358 (2000).
\bibitem{Re97a} M. P. Rekalo, J. Arvieux and E. Tomasi-Gustafsson, Z. Phys. Rev. A {\bf 357}, 133 (1997).
\bibitem{Re97b} M. P. Rekalo, J. Arvieux and E. Tomasi-Gustafsson, Phys. Rev. C {\bf 55}, 1630 (1997).
\bibitem{Re97c} M. P. Rekalo, J. Arvieux and E. Tomasi-Gustafsson, Phys. Rev. C {\bf 56}, 2238 (1997).
\bibitem{El95} J. Ellis, M. Karliner, D. E. Kharzeev, and M. G. Sapozhnikov,
Phys. Lett. {\bf B353} 319 (1995). 
\bibitem{Pak99} N. K. Pak and M. P. Rekalo, Phys. Rev. C {\bf 59}, 077501 (1999).
\bibitem{Fa90} G. F\"aldt and C. Wilkin, Nucl. Phys. {\bf A518}, 308 (1990).
\bibitem{La91} J.~M.~Laget, F. Wellers and J.F. Lecolley, Phys. Lett. 
{\bf B257}, 254 (1991).  
\bibitem{Ve91} T. Vetter A.~Engel, T.~Biro and U.~Mosel, Phys. Lett. {\bf B263}, 153 (1991).

\bibitem{Gr00}
V.~Y.~Grishina, L.~A.~Kondratyuk, M.~Buscher, C.~Hanhart, J.~Haidenbauer and J.~Speth,
Phys.\ Lett.\ B {\bf 475}, 9 (2000).
\bibitem{Ta00} B. Tatischeff {\it et al.}, Phys. Rev. {\bf C62}, 054001 (2000); 
\bibitem{Kl80}
W.~M.~Kloet and R.~R.~Silbar,
Nucl.\ Phys.\ A {\bf 338}, 281 (1980);\\
Nucl.\ Phys.\ A {\bf 364}, 346 (1981).

\bibitem{Mi93}
T.~Mizutani, C.~Fayard, G.~H.~Lamot and B.~Saghai,
Phys.\ Rev.\ C {\bf 47}, 56 (1993).
\bibitem{Au89}
J.P. Auger, C. Lazard and R. J. Lombard, Phys. Rev.  D {\bf 39}, 56 (1989).
\bibitem{Ha94} S. Haber and J. Aichelin, Nucl. Phys. A {\bf 573}, 587 (1994).
\bibitem{Yo98}
J. Yonnet {\it et al.}, Nucl. Phys.  A {\bf 637}, 63 (1998).
\bibitem{Po53} I. Ya. Pomeranchuk and E. L. Feinberg, Dokl. Akad. Nauka Ser. Fiz. 93, 439, (1953). 

\bibitem{Do83} A. Donnachie and L. V. Landshoff,  Phys. Lett. B {\bf 123}, 345 (1983);\\
Nucl. Phys. B {\bf 231}, 189 (1984). 

\end{thebibliography}
\end{document}